\documentclass[twocolumn,10pt,prb,groupedaddress,showpacs,preprintnumbers,amsmath,amssymb,twoside]{revtex4}

\usepackage{graphicx}
\usepackage{dcolumn}
\usepackage{bm}

\begin{document}
\preprint{Submitted to Phys. Rev. B}

\title{Vibrational exciton-mediated quantum state transfert: a simple model.} 

\author{Vincent Pouthier}
\email{vincent.pouthier@univ-fcomte.fr}
\affiliation{Institut UTINAM,  Universit\'{e} de Franche-Comt\'{e}, \\  CNRS UMR 6213, 25030 Besan\c {c}on Cedex, 
France}

\date{\today}

\begin{abstract}
A communication protocol is proposed in which quantum state transfer is mediated by a vibrational exciton. We consider two distant molecular groups grafted on the sides of a lattice. These groups behave as two quantum computers where the information in encoded and received. The lattice plays the role of a communication channel along which the exciton propagates and interacts with a phonon bath. Special attention is paid for describing the system involving an exciton dressed by a single phonon mode.
The Hamiltonian is thus solved exactly so that the relevance of the perturbation theory is checked. 
Within the nonadiabatic weak-coupling limit, it is shown that the system supports three quasi-degenerate states that define the relevant paths followed by the exciton to tunnel between the computers. When the model parameters are judiciously chosen, constructive interferences take place between these paths. Phonon-induced decoherence is minimized and a high-fidelity quantum state transfer occurs over a broad temperature range. 
\end{abstract}

\maketitle

\section{Introduction}

Over short length scales, high-fidelity quantum-state transfer (QST) from one region to another is a fundamental task in quantum information processing \cite{kn:bennet}. It is required to ensure a perfect communication between the different parts of a quantum computer (QC) or  between adjacent QCs. In that case, solid-state  based system  is the ideal candidate for scalable computing for, at least, two main reasons \cite{kn:burgarth3}. First, in a lattice,  the information is encoded on elementary excitations that naturally propagate owing to inherent interactions between neighboring sites. QST is thus spontaneously achieved without any external control. Second, no interfacing is needed between the QCs and the communication channel (CC) that involves the same elementary excitations. Because the CC depends on the way the information is implemented, different strategies have been elaborated, such as optical lattices\cite{kn:mandel}, arrays of quantum dots\cite{kn:loss}, conducting polymers\cite{kn:huo} and spin networks\cite{kn:bose,kn:christandl1,kn:christandl2,kn:li,kn:karbach,kn:burgarth1,kn:yung,kn:burgarth2,kn:hu,kn:cappellaro,kn:yao}. However, it has been shown that qubits may be encoded on intramolecular vibrations \cite{kn:riedle1,kn:riedle2,kn:riedle3,kn:riedle4,kn:riedle5,kn:riedle6,kn:riedle7} so that vibration-mediated QST is a promising alternative for quantum information processing \cite{kn:plenio,kn:gollub}.

In this paper, we propose a protocol in which a molecular lattice plays the role of the CC, QST being mediated by high-frequency vibrational excitons. Indeed, molecular lattices exhibit regularly distributed atomic subunits. Owing to dipole-dipole interactions, the energy of a specific internal vibration delocalizes between these subunits giving rise to narrow-band excitons. 
Although many properties of the excitons have been studied\cite{kn:davydov,kn:scott,kn:PRE03,kn:edler,kn:falvo,kn:tsivlin,kn:cruzeiro,kn:JCP08,kn:PRE08,kn:JPC09,kn:bittner10,kn:persson,kn:PRB99,kn:bonn1,kn:JCP01,kn:jakob,kn:PRB02,kn:JCP03,kn:bonn2,kn:PRB05,kn:PRB06}, their  potential interest for QST has been suggested very recently\cite{kn:PRB2011a,kn:JCP2011,kn:PRB2011b}. The main idea is to encode the information on a vibrational qubit defined as a superimposition involving the vacuum and one-exciton states. 

Unfortunately, the exciton does not propagate freely along the lattice. It interacts with a phonon bath that tends to destroy the coherent nature of any vibrational qubit. The exciton properties are thus described by a reduced density matrix (RDM) whose behavior is governed by a generalized master equation (GME)\cite{kn:book0,kn:book1,kn:book2}. To understand quantum decoherence, we have studied excitonic coherences, i.e. the RDM matrix elements that measure the ability of the exciton to be in a qubit state\cite{kn:JCP10,kn:JPC10a,kn:PRE10,kn:JPC10b}. The dynamics was described using a Fr\"{o}hlich model\cite{kn:frohlich} within the nonadiabatic weak-coupling limit, i.e. a common situation for vibrational excitons. It has been shown that high-fidelity QST requires the use of a finite-size system. 

Indeed, in an infinite lattice\cite{kn:JCP10,kn:JPC10a}, the phonons behave as a reservoir. The Markov limit is reached and standard GME approaches can be used. Dephasing limited-coherent motion takes place so that the coherences localize preventing any efficient QST. By contrast, in a finite-size lattice, a strong non-markovian regime occurs resulting in the breakdown of GME methods\cite{kn:PRE10,kn:JPC10b}. To overcome this problem, perturbation theory (PT) has been applied successfully\cite{kn:PRB2011a,kn:JCP2011,kn:PRB2011b}. It has been shown that the phonons evolve differently depending on whether the exciton occupies the vacuum or an excited state. Exciton-phonon entanglement takes place resulting in the decay of the excitonic coherences. Nevertheless, the decoherence depends on the nature of the one-exciton state. 

In particular, for odd lattice sizes, the coherence involving the state located at the band center survives over an extremely long-time scale, even at high temperature. Unfortunately, this state is a stationary wave unable to carry information between the lattice sides. Nevertheless, we can take advantage of its robustness against decoherence to define a protocol quite similar to that introduced by Yao et al. in spin chains\cite{kn:yao}. 
The main idea is to consider QST between two distant molecular groups that are grafted on the lattice sides. 
The structure must be designed so that a vibration of each molecular group is resonant with the robust state of the lattice while remaining insensitive to the phonons. The grafted groups play the role of the QCs whereas the lattice defines the CC. Dipole-dipole interactions couple the QCs with the CC and allow the exciton propagation so that neither control nor interfacing is required. 

In such a confined system, strong non-markovian effects will occur so that the dynamics will be addressed within the effective Hamiltonian concept provided by PT\cite{kn:wagner,kn:cohen}. However, the fundamental question arises whether PT is relevant or not, depending on the model parameters. To answer that question, we shall restrict our attention to a system in which only the lowest frequency phonon mode (LFPM) is considered, this mode being responsible for the largest perturbation of the exciton dynamics\cite{kn:PRB2011a,kn:JCP2011}. Due to its simplicity, this model can be solved exactly so that the PT performance can been checked easily. 

The paper is organized as follows. In Sec. II, the exciton-phonon Hamiltonian is described and QST is formulated in terms of the excitonic coherences. Then, PT is applied to evaluate the system eigenproperties and to determine an approximate expression of the exciton RDM. In Sec. III, a numerical analysis is performed in which PT is compared with exact calculations. The results are discussed and interpreted in Sec. IV.

\begin{figure}
\includegraphics{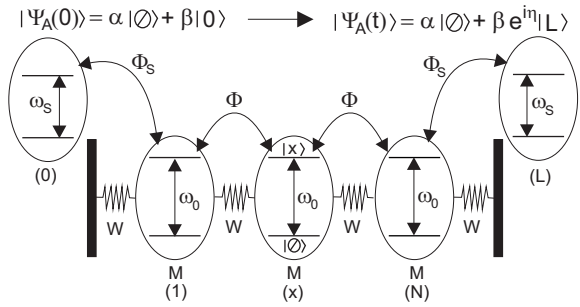}
\caption{Communication protocol between two distant computers $x=0$ and $x=L$. The information is carried by an exciton that propagates along a finite-size lattice.}
\end{figure}

\section{Theoretical Background}

\subsection{Model Hamiltonian}

We consider a CC formed by a 1D molecular lattice that contains an odd number of sites $N$ (Fig. 1). Each site $x=1,...,N$ is occupied by an atomic subunit whose internal dynamics is described by a two-level system. Let $|x\rangle$ be the first excited state of the $x$th two-level system and $\omega_0$ the corresponding Bohr frequency. The vacuum state $|\oslash_{cc} \rangle$ describes all the two-level systems in their ground state. In the CC, the exciton Hamiltonian is defined in terms of the hopping constant $\Phi$ as ($\hbar=1$)
\begin{equation} 
H_{cc}= \sum_{x=1}^{N} \omega_0 |x\rangle \langle x|+\sum_{x=1}^{N-1} \Phi (|x+1\rangle \langle x|+|x\rangle \langle x+1|).
\label{eq:Hcc}
\end{equation}
Owing to the confinement, one-exciton states are $N$ stationary waves with quantized wave vectors $K_{k}=k\pi/L$, with $k=1,..,N$ and $L=N+1$, expressed as 
\begin{equation} 
|\varphi_k\rangle= \sum_{x=1}^{N}  \sqrt{\frac{2}{L}} \sin( K_k x) |x\rangle.
\label{eq:ketk}
\end{equation}
The corresponding eigenenergies $\omega^0_{k}=\omega_0+2\Phi \cos(K_k)$ form $N$ discrete energy levels centered on $\omega_0$, i.e. the energy of the so-called robust stationary wave $|\varphi_{L/2}\rangle$. 
The phonons of the CC refer to the external motions of the lattice sites that behave as point masses $M$ connected via force constants $W$. They define $N$ normal modes with wave vectors $Q_p=p\pi/L$ and frequencies $\Omega_{p}=\Omega_{c} \sin(Q_p/2)$, with $p=1,..,N$ and $\Omega_{c}=\sqrt{4W/M}$. Restricting our attention to the LFPM $p=1$, the phonon Hamiltonian is defined as $H_{B}=\Omega a^{\dag}a$, with $\Omega\equiv\Omega_{1}$, $a^{\dag}$ and $a$ being standard boson operators. In the phonon Hilbert space $\mathcal{E}_B$, the eigenstates are thus the well-known number states $|n\rangle$.
As detailed previously\cite{kn:PRB2011a}, the exciton-phonon interaction $V=M(a^{\dag}+a)$ favors the exciton scattering from $|\varphi_k\rangle$ to $|\varphi_{k\pm1}\rangle$ mediated by phonon exchanges. The matrix elements of the $M$ operator are  defined as 
\begin{equation}
M_{kk'} = \eta (\delta_{k,k'+1}+\delta_{k,k'-1}),
\label{eq:V}
\end{equation}
where $\eta = [(E_B \Omega /L) ( 1- (\Omega/\Omega_c)^2 )]^{1/2}$, $E_B$ being the small polaron binding energy expressed in terms of the coupling strength $\chi$ as $E_B=\chi^2/W$.

The QCs are formed by two molecular groups $x=0$ and $x=L$ 
whose internal dynamics is described by a two-level system (Fig. 1). Let $|0\rangle$ (resp. $|L\rangle$) be the first excited state of the $0$th (resp. $L$th) molecular group and $\omega_S$ the corresponding Bohr frequency. The vacuum state $|\oslash_{qc}\rangle$ describes the two groups in their ground state. Located far enough from each other, these groups
do not interact. Their internal dynamics is governed by the Hamiltonian $H_{qc}=\omega_S (|0\rangle \langle 0|+|L\rangle \langle L|)$. 

The QCs, assumed to be sufficiently far from the lattice, are insensitive to the phonons. By contrast, owing to dipole-dipole interactions, they interact with the lattice exciton. This interaction originates in a vibrational energy transfer between the group $0$ (resp. $L$) and the lattice site $x=1$ (resp. $x=N$) as
\begin{equation} 
W=\Phi_S (|0\rangle \langle 1|+ |1\rangle \langle 0|+|L\rangle \langle N|+ |N\rangle \langle L|),
\label{eq:W}
\end{equation}
where $\Phi_S=\epsilon \Phi$, $\epsilon$ being the strength of the coupling between the QCs and the CC.  

Within this model, the exciton dynamics is governed by the Hamiltonian $H_A=H_{cc}+H_{qc}+W$. It acts in the Hilbert space $\mathcal{E}_A$ whose dimension is $N+3$. $\mathcal{E}_A$ is generated by the vacuum $|\oslash \rangle=|\oslash_{cc} \rangle \otimes |\oslash_{qc} \rangle$ and by $N+2$ one-exciton eigenstates $|\psi_{\mu}\rangle$, associated to the  eigenfrequencies $\omega_{\mu}$, with $\mu=0,...,L$. 
\begin{figure}
\includegraphics{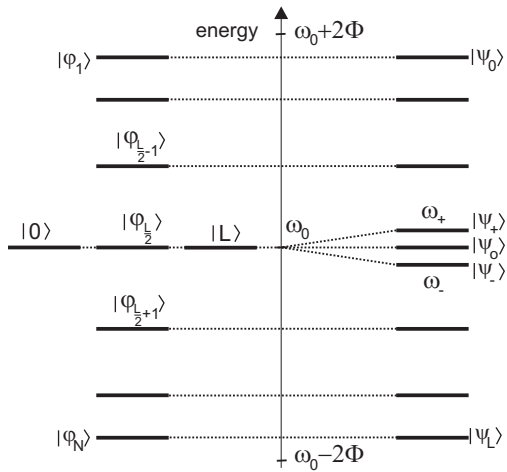}
\caption{Exciton energy spectrum for $\epsilon=0$ (left side) and for $\epsilon\neq0$ (right side).}
\end{figure}
The exciton Hamiltonian is thus rewritten as  $H_A=\sum_{\mu=0}^L\omega_{\mu} |\psi_{\mu}\rangle \langle \psi_{\mu}|$. 
To determine these one-exciton states, we first fix $\omega_S$ to $\omega_0$ so that a resonance occurs between the two localized states $|0\rangle$ and $|L\rangle$ and the robust stationary wave $|\varphi_{L/2}\rangle$ (Fig. 2). Then, 
$\epsilon$ is assumed to be sufficiently small so that off-resonant interactions between the localized states and the remaining stationary waves can be disregarded. This condition is satisfied provided that $\epsilon \ll \pi \sqrt{2/L}$. As a result, the one-exciton eigenstates $|\psi_{\mu}\rangle$ are split into two groups (Fig. 2).

The first group contains $N-1$ states that reduce to the stationary waves $|\varphi_{k}\rangle$, with $k\neq L/2$. 
For $\mu=0,...,L/2-2$, they are defined as $|\psi_{\mu}\rangle=|\varphi_{\mu+1}\rangle$ ($\omega_{\mu}=\omega_{\mu+1}^0$) whereas for $\mu=L/2+2,...,L$, they reduce to $|\psi_{\mu}\rangle=|\varphi_{\mu-1}\rangle$ ($\omega_{\mu}=\omega_{\mu-1}^0$). 
The second group results from the hybridization between the degenerate states 
$|0\rangle$, $|\varphi_{L/2}\rangle$ and $|L\rangle$. Indeed, $|0\rangle$ (resp. $|L\rangle$) interacts with $|\varphi_{L/2}\rangle$ through the coupling constant $g=\epsilon \Phi\sqrt{2/L}$ (resp. $g'=g \sin( N\pi/2)$). When $\epsilon\neq 0$, these couplings raise the degeneracy so that the eigenstates are defined as 
\begin{eqnarray}
|\psi_{\pm}\rangle&=&\frac{1}{2}|0\rangle \pm \frac{1}{\sqrt{2}}|\varphi_{L/2}\rangle +\frac{\Delta_N}{2}|L\rangle \nonumber \\
|\psi_o\rangle&=&\frac{1}{\sqrt{2}}|0\rangle-\frac{\Delta_N}{\sqrt{2}}|L\rangle,
\label{eq:STATE1}
\end{eqnarray}
where $\Delta_N =\sin(N\pi/2)$ and with the convention $\mu=L/2\mp1\equiv \pm$ and $\mu=L/2\equiv o$. The corresponding eigenfrequencies are expressed as
\begin{eqnarray}
\omega_{\pm}&=&\omega_0\pm2\epsilon \Phi/\sqrt{L} \nonumber \\
\omega_o&=&\omega_0.
\label{eq:STATE2}
\end{eqnarray}
$|\psi_{\pm}\rangle$ and $|\psi_{o}\rangle$ define three quasi-degenerate states that involve a superimposition of the localized states, either symmetric or antisymmetric. Note that $|\psi_{o}\rangle$ is exactly located at the band center and it does not depend on the stationary wave $|\varphi_{L/2}\rangle$. This is no longer the case for $|\psi_{+}\rangle$ and $|\psi_{-}\rangle$ that lie just above and just below the band center, respectively. 

The exciton-phonon Hamiltonian is thus written as $H=H_0+V$ where $H_0=H_A+H_B$ is the unperturbed Hamiltonian. It acts in the Hilbert space $\mathcal{E}=\mathcal{E}_A\otimes \mathcal{E}_B$ that is partitioned into independent subspaces as $\mathcal{E}=\mathcal{E}_0\oplus \mathcal{E}_1$. In the zero-exciton subspace $\mathcal{E}_0$, $V=0$ so that the unperturbed states are eigenstates of $H$. They correspond to tensor products $|\oslash \rangle \otimes | n\rangle$ that describe $n$ free phonons. In the one-exciton subspace $\mathcal{E}_1$, the unperturbed states $|\Psi_{\mu,n}^0\rangle=|\psi_{\mu} \rangle \otimes | n \rangle$ refer to free phonons accompanied by an exciton in state $|\psi_{\mu}\rangle$. Because $V$ turns on, they are no longer eigenstates of $H$. The exact eigenstates $|\Psi_{i} \rangle$, with eigenenergies $E_i$, correspond to entangled exciton-phonon states. 

The coupling $V$ favors exciton scattering from $|\psi_{\mu}\rangle$ to $|\psi_{\mu'}\rangle$ through phonon exchanges. The allowed transitions are specified by the selection rules $\langle \psi_{\mu}| M|\psi_{\mu'}\rangle\neq0$ (Eq.(\ref{eq:V})). Because $|\psi_{\mu}\rangle$ interacts with the phonons through its dependence with respect to the stationary waves, these rules are quite similar to those that characterize the CC ($|\psi_{\mu}\rangle$ is coupled with $|\psi_{\mu\pm1}\rangle$). However, two main differences occur. First, because $|\psi_{o}\rangle$ only depends on the localized states, it remains insensitive to the phonon bath. Second, a splitting occurs for the transitions involving $|\psi_{\pm}\rangle$. Indeed, because $|\psi_{L/2\pm2}\rangle$ reduces to $|\varphi_{L/2\pm1}\rangle$, it interacts with $|\psi_{+}\rangle$ and $|\psi_{-}\rangle$, both states depending on $|\varphi_{L/2}\rangle$. 
Nevertheless, within the nonadiabatic limit, i.e. for $4\Phi < \Omega_c$ and $\epsilon \ll \pi \sqrt{2/L}$, the allowed transitions do not conserve the energy. There is no resonance between the coupled unperturbed states so that 
second order PT can be applied to treat $V$ in the weak coupling limit ($E_B \ll \Phi $). As detailed previously\cite{kn:PRB2011a,kn:JCP2011}, PT is valid at temperature $T$ provided that $L<L^{*}$ with $L^{*}\approx0.2\Omega_c^2/E_B k_BT$. When $L>L^{*}$, quasi-resonances take place between the unperturbed coupled states and PT breaks down.

\subsection{Excitonic coherences and QST}

Without any perturbation, the CC and the QCs are in thermal equilibrium at temperature $T$. Assuming that $\omega_0\gg k_BT$, all the two-level systems are in their ground state. By contrast, the phonons form a thermal bath described by the density matrix $\rho_B=\exp(-\beta H_B)/Z_B$, $Z_B$ being the phonon partition function ($\beta=1/k_BT$). 
In that case, one assumes that the internal vibrations interact with an external source so that the exciton is initially prepared in a state $|\psi_A\rangle \neq |\oslash\rangle$. This step is supposed to be rather fast when compared with the typical time evolution of the phonons. The full system is thus brought in a configuration out of equilibrium and its initial density matrix is defined as $\rho(0)=|\psi_A\rangle \langle \psi_A | \otimes \rho_B$. 

To study QST between the QCs, $|\psi_A\rangle$ defines a qubit implemented on the molecular group $x=0$ as
\begin{equation}
|\psi_A\rangle = \alpha |\oslash\rangle + \beta |0\rangle,
\end{equation}
where $|\alpha|^2+|\beta|^2=1$. Our aim is thus to measure the ability of the system to freely evolve in time so that this initial qubit is copied on the second QC $x=L$ (Fig. 1). Whatever its duration, QST must be realized with the largest fidelity despite the coupling with the phonons. To define this fidelity measure, different objects have been introduced\cite{kn:fidelity}, one of the most widely used being certainly the so-called average Schumacher's fidelity\cite{kn:bose}. Here, we restrict our attention to the excitonic coherences. Indeed, the exciton properties are encoded in the RDM $\sigma(t)=Tr_B [ \rho(t)]$, where $Tr_B$ is a partial trace over the phonon degrees of freedom. The coherences are thus the off-diagonal matrix elements $\sigma_{x\oslash}(t)$. They provide information about the ability of the $x$th two-level system to develop a superimposition between its ground state and its first excited state a time $t$.  

When the initial qubit is implemented on the $0$th QC, the coherence $\sigma_{0\oslash}(0)$ is turned on at time $t=0$. Therefore, the ability of the $L$th QC to develop a superimposition involving $|\oslash\rangle$ and $|L\rangle$ at time $t$ is given by $\sigma_{L\oslash}(t)= G_{L0}(t)\sigma_{0\oslash}(0)$ with   
\begin{equation}
G_{L0}(t) =\langle L | Tr_B \left[ \rho_B e^{iH_Bt} e^{-iHt} \right] |0\rangle.
\label{eq:GL0}
\end{equation}
The effective exciton propagator $G_{L0}(t)$ generalizes the concept of transition amplitude. It yields the probability amplitude to observe the exciton in $|L\rangle$ at time $t$ given that it was in $|0\rangle$ at $t=0$. Its effective nature results from the fact that the exciton interacts with the phonons during its transition. 

The effective propagator is the central object of the present study. The condition $|G_{L0}(t)|= 1$ reveals that the $L$th QC reaches a state at time $t$ that is equivalent to the initial state, to a phase factor. Note that this condition  is exactly the QST fidelity when the coupling with the phonons is disregarded\cite{kn:bose}. Consequently, depending on the value of the model parameters, studying the maximum value of $|G_{L0}(t)|$ provides key information about the fidelity of the QST. 

\subsection{Perturbation theory}

In its operatorial formulation\cite{kn:wagner}, standard PT involves a unitary transformation that provides a new point of view in which the exciton-phonon dynamics is described by an effective Hamiltonian. The key point is that this Hamiltonian is diagonal in the unperturbed basis. Quite powerful to treat finite-size lattices\cite{kn:PRB2011a,kn:JCP2011}, this approach breaks down in the present situation because the unperturbed Hamiltonian exhibits quasi-degenerate states. For small 
$\epsilon$ values, $|\Psi_{+,n}^{0}\rangle$ and $|\Psi_{-,n}^{0}\rangle$ have almost the same energy. Although they do not directly interact through $V$, they are coupled with the same unperturbed states. Consequently, effective couplings occur between quasi-degenerate states resulting in errors in the calculations of the corrected energies.

To overcome this problem, quasi-degenerate PT is applied\cite{kn:cohen} (Appendix A). To proceed, we take advantage of the fact that the effective couplings conserve the phonon number. Therefore, our procedure involves a transformation $U=\exp(S)$ that generates a new point of view in which the effective Hamiltonian $\hat{H}=UHU^{\dag}$ is block-diagonal in the unperturbed basis. The generator $S$ is expanded as a Taylor series in the coupling $V$ so that $\hat{H}$ becomes diagonal with respect to the phonon number states, only.
Up to second order, it is written as
\begin{equation}
\hat{H}=H_A+\delta H_A + (\Omega + \delta \Omega) a^{\dag} a
\label{eq:HEFF}
\end{equation}
where $\delta H_A$ and $\delta \Omega$ are operators in $\mathcal{E}_A$ whose matrix elements are defined as 
(in the unperturbed basis $\{ |\psi_{\mu} \rangle \}$)
\begin{eqnarray}
\delta H_{A\mu_1 \mu_2}&=&\frac{1}{2} \sum_{\mu=0}^{L} \frac{M_{\mu_2\mu}M_{\mu \mu_1}}{\omega_{\mu_1}-\omega_{\mu}-\Omega}+
\frac{M_{\mu_1\mu}M_{\mu \mu_2}}{\omega_{\mu_2}-\omega_{\mu}-\Omega} \nonumber \\
\delta \Omega_{\mu_1 \mu_2}&=&\frac{1}{2} \sum_{\mu=0}^{L} \frac{M_{\mu_2\mu}M_{\mu \mu_1}}{\omega_{\mu_1}-\omega_{\mu}-\Omega}+
\frac{M_{\mu_1\mu}M_{\mu \mu_2}}{\omega_{\mu_2}-\omega_{\mu}-\Omega} \nonumber \\
&+&\frac{1}{2} \sum_{\mu=0}^{L} \frac{M_{\mu_2\mu}M_{\mu \mu_1}}{\omega_{\mu_1}-\omega_{\mu}+\Omega}+
\frac{M_{\mu_1\mu}M_{\mu \mu_2}}{\omega_{\mu_2}-\omega_{\mu}+\Omega}
\label{eq:deltaH}
\end{eqnarray}

$\delta H_A$ is the correction of the exciton Hamiltonian owing to the coupling with the phonons. It results from the spontaneous emission of a phonon during which the exciton realizes a transition from $|\psi_{\mu_1}\rangle$ to $|\psi_{\mu}\rangle$. However, in the nonadiabatic limit, the energy is not conserved during the transition. The emitted phonon is immediately reabsorbed and the exciton realizes a second transition from $|\psi_{\mu}\rangle$ to $|\psi_{\mu_2}\rangle$. In other words the exciton does no longer propagate freely but it is dressed by a virtual phonon cloud. This dressing renormalizes the excitonic energies $\omega_{\mu}$ by an amount $\delta \omega_{\mu}=\delta H_{A\mu\mu}$. In addition, it induces effective interactions $\delta H_{A\mu_1 \mu_2}$ between distinct excitonic states that can no longer be neglected for quasi-degenerate states.

Similarly,  $\delta \Omega$ defines the correction of the phonon frequency. It has two origins. First, the phonon can be absorbed giving rise to excitonic transitions. Because this process does not conserve the energy, the phonon is  immediately re-emitted. Second, the phonon can induce the stimulated emission of a second phonon during which the exciton realizes transitions. But, as previously, the emitted phonon is immediately reabsorbed. Both mechanisms are virtual processes indicating that the phonons are dressed by virtual excitonic transitions.

To diagonalize $\hat{H}$ for each phonon number, we use the fact that $\delta \Omega$ is smaller than $\delta H_A$ within the weak coupling limit. Consequently, let $|\chi_{\nu}\rangle$ be the eigenstates of $H_A+\delta H_A$ and $\hat{\omega}_{\nu}$ the corresponding eigenfrequencies ($\nu=0,...,L$). Up to second order in $V$, one introduces $\delta \Omega_{\nu}= \langle \chi_{\nu}| \delta \Omega  |\chi_{\nu}\rangle$ as the correction of the phonon frequency induced by the exciton that occupies the state $|\chi_{\nu}\rangle$. Within these notations, $\hat{H}$ is rewritten as 
\begin{equation}
\hat{H}\approx \sum_{\nu=0}^{L}  \hat{\omega}_{\nu} |\chi_{\nu}\rangle\langle \chi_{\nu}|+\hat{H}_B^{(\nu)}\otimes|\chi_{\nu}\rangle\langle \chi_{\nu}|,
\label{eq:HEFF2}
\end{equation}
where $\hat{H}_B^{(\nu)}=(\Omega+\delta \Omega_{\nu})a^{\dag}a$ is the Hamiltonian that governs the phonon dynamics when the exciton is in the state $|\chi_{\nu}\rangle$. 

In the new point of view, the exciton-phonon dynamics is thus governed by the effective Hamiltonian $\hat{H}$ that is diagonal in the basis $|\chi_{\nu}\rangle\otimes|n\rangle$. Its eigenvalues define the system eigenfrequencies up to second order in $V$ as $E_{\nu,n}= \hat{\omega}_{\nu}+n(\Omega+\delta \Omega_{\nu})$. $\hat{H}$ does no longer refer to independent excitations but it characterizes entangled exciton-phonon states. A state $|\chi_{\nu}\rangle$ describes an exciton dressed by a virtual phonon cloud whereas the number state $| n \rangle$ describes $n$ phonons clothed by virtual excitonic transitions. This entanglement is clearly highlighted in the starting point of view in which the eigenstates do no longer factorize as $|\Psi_{\nu,n} \rangle = U^{\dag} |\chi_{\nu}\rangle \otimes | n \rangle$. 

As detailed previously\cite{kn:JCP2011}, PT is particularly suitable for deriving an approximate expression for $G_{L0}(t)$. To proceed, we first introduce $U$ and diagonalize $H$ in Eq.(\ref{eq:GL0}). Then, we define the effective density matrix 
\begin{equation}
\rho_B^{(\nu)}(t)=\frac{1}{Z_B^{(\nu)}(t)} \exp \left[ - (\beta \Omega+i\delta \Omega_{\nu}t)a^{\dag} a \right],
\label{eq:rhoBnu}
\end{equation}
where $Z_B^{(\nu)}(t)=(1-\exp \left[ -(\beta \Omega+i\delta \Omega_{\nu}t)\right])^{-1}$.
Strictly speaking, $\rho_B^{(\nu)}(t)$ is not a density matrix since it yields complex values for the phonon population. However, it is isomorphic to $\rho_B$ with the correspondence $\beta \Omega \rightarrow  \beta \Omega+i\delta \Omega_{\nu}t$  and it provides averages equivalent to thermal averages. Consequently, after simple algebraic manipulations, $G_{L0}(t)$ is rewritten as\cite{kn:JCP2011}
\begin{eqnarray}
G_{L0}(t) &=&\sum_{\nu=0}^L \frac{Z_B^{(\nu)}(t)}{Z_B} \exp[-i\hat{\omega}_{\nu}t] \times \label{eq:GL01}    \\
&&Tr_B \left[ \rho_B^{(\nu)}(t) \langle L |U^{\dag}_{\nu}(t)|\chi_{\nu} \rangle \langle \chi_{\nu} | U_{\nu}(0)|0\rangle \right]. \nonumber
\end{eqnarray}
where $U_{\nu}(t)=e^{i\hat{H}_B^{(\nu)} t} U e^{-i\hat{H}_B^{(\nu)} t}$.
Expanding $U$ in a Taylor series with respect to $V$, one finally obtains the second order expression of $G_{L0}(t)$, as detailed in Appendix B. 

\section{Numerical results}

In this section, numerical calculations are carried out to show the relevance of PT for describing QST. To proceed, the previous formalism is applied to amide-I vibrations in $\alpha$-helices, a system for which the parameters are well-known\cite{kn:davydov,kn:scott,kn:PRE03,kn:edler,kn:falvo,kn:tsivlin,kn:cruzeiro,kn:JCP08,kn:PRE08,kn:JPC09,kn:bittner10} : $\omega_0=1660$ cm$^{-1}$, $W=15$ Nm$^{-1}$, $M=1.8\times 10^{-25}$ kg, $\Omega_c=96.86$ cm$^{-1}$ and $\Phi=7.8$ cm$^{-1}$. To avoid PT breakdown, the size is set to $L=10$ and the exciton-phonon coupling strength is fixed to $\chi=10$ pN. Special attention will be paid for characterizing the influence of the coupling $\epsilon$ between the QCs and the CC. 

\begin{figure}
\includegraphics{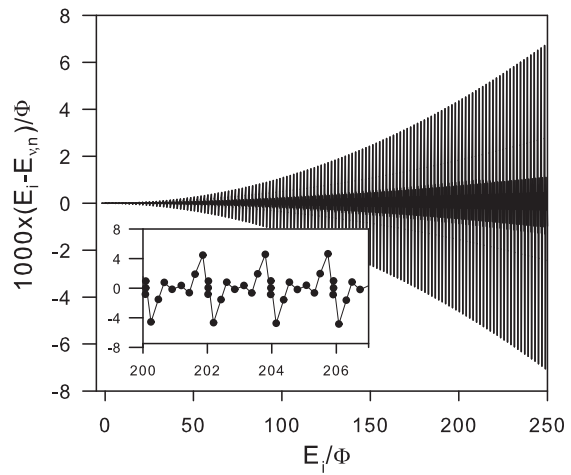}
\caption{$E_i-E_{\nu,n}$ vs $E_i$ for $\epsilon=0.01$, $\chi=10$ pN and $L=10$. }
\end{figure}

The difference $\Delta E= E_{i}-E_{\nu,n}$ between exact and approximate energies is shown in Fig. 3. For $\epsilon=0.01$, PT provides a very good estimate of the energy spectrum over a broad energy scale. The smaller the energy is, the better is the agreement. Of course, $\Delta E$ increases with $E_{i}$ indicating that the PT accuracy decreases with the phonon number $n$ because $V$ scales as $\sqrt{n}$. For instance, for $E_i\approx100\Phi$ ($n\approx50$), $\Delta E$ is approximately equal to $10^{-3}\Phi$ whereas it reaches $5 \times 10^{-3}\Phi$ for $E_i\approx200\Phi$.
Nevertheless, we have verified that $\Delta E$ is smaller than the energy level spacing indicating that PT remains valid, even for quite large energies. As shown in the inset, the curve $\Delta E$ vs $E_{i}$ behaves almost periodically, with a period approximately equal to $\Omega$, indicating that the PT accuracy depends on the nature of the unperturbed states. PT is exact for the unperturbed states that involve the exciton state insensitive to the phonons. By contrast, unperturbed states located in the neighborhood of quasi-degenerate states are the less well-corrected. 

\begin{figure}
\includegraphics{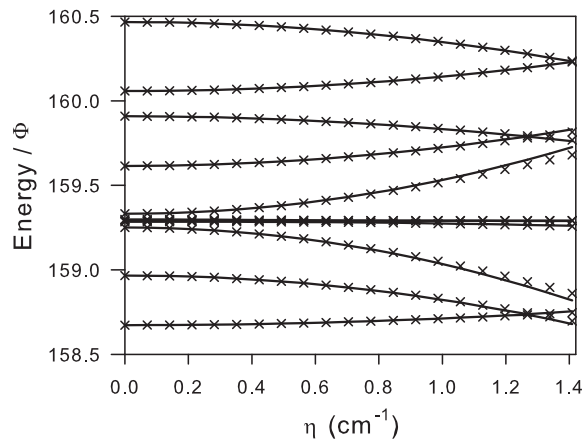}
\caption{$E_i$ (thin x) and $E_{\nu,n}$ (full lines) vs $\eta$ for $\epsilon=0.01$ and $L=10$.}
\end{figure}

As shown in Fig. 4, $\Delta E$ slightly increases with $\chi$ so that PT remains valid in the intermediate coupling regime. It reproduces the influence of the exciton-phonon coupling on the energy levels up to $\chi=20$ pN ($\eta=1.41$ cm$^{-1}$). In particular, PT accounts for the energy level crossing process that affects the dressed states. Owing to the coupling $V$, unperturbed states hybridize giving rise to energy shift and splitting characteristic of anti-crossing phenomena. Some energy levels repel each other whereas other energy levels get closer. These levels describing exact uncoupled eigenstates, energy level crossing takes place. 

\begin{figure}
\includegraphics{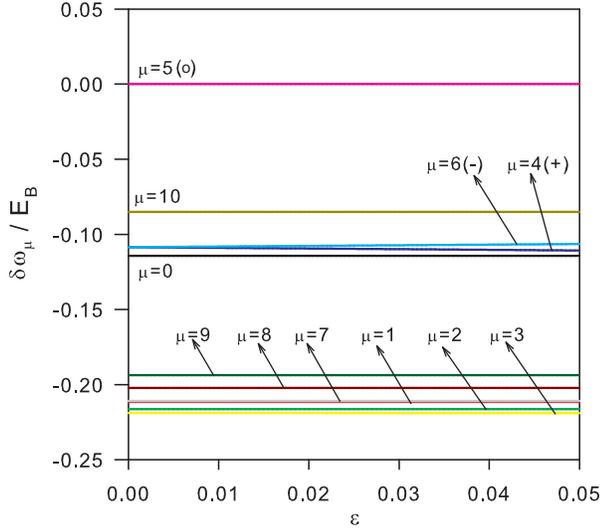}
\caption{(Color online) $\delta \omega_{\mu}$ vs $\epsilon$ for $\chi=10$ pN and $L=10$.}
\end{figure}

The $\epsilon$ dependence of the exciton energy correction $\delta \omega_{\mu}$ is shown in Fig. 5. Owing to the dressing by virtual phonons, the exciton energy is redshifted and $\delta \omega_{\mu}$ decreases linearly with $E_B$. However, this behavior depends on the exciton state. For the state $|\psi_{o}\rangle $ insensitive to the phonons, $\delta \omega_{o}=0$. By contrast, the stationary waves $\mu \neq o,\pm $ experience an energy correction approximately equal to $-0.2E_B$. This shift reduces to $-0.1E_B$ for the states $|\psi_{0}\rangle$ and $|\psi_{L}\rangle $, whose energies define the band edges, and for the quasi-degenerate states $|\psi_{\pm}\rangle$. Moreover, Fig. 5 reveals that the energy correction of the stationary waves is almost $\epsilon$ independent. This is no longer the case for the quasi-degenerate states because $\delta \omega_{+}$ (resp. $\delta \omega_{-}$) decreases (resp. increases) linearly with $\epsilon$ (not well distinguishable in Fig. 5). Note that $\delta \omega_{+}=\delta \omega_{-}=0.108E_B$ when $\epsilon=0$.

\begin{figure}
\includegraphics{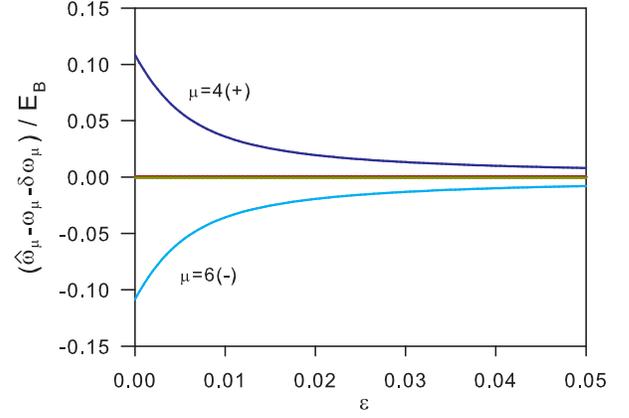}
\caption{(Color online) $\hat{\omega}_{\mu}-\omega_{\mu}-\delta \omega_{\mu}$ vs $\epsilon$ for $\chi=10$ pN and $L=10$.}
\end{figure}

Fig. 6 shows the $\epsilon$ dependence of the difference between the energy $\hat{\omega}_{\mu}$ of a dressed state $|\chi_{\mu}\rangle$ and the corrected energy $\omega_{\mu}+\delta \omega_{\mu}$ of a bare state $|\psi_{\mu}\rangle$. For the stationary waves $\mu\neq o,\pm$, a correspondence occurs between both energies suggesting that the dressing mainly induces an energy renormalization without significantly modifying the state. In fact, we have verified that  $|\chi_{\mu}\rangle\approx |\psi_{\mu}\rangle$ for $\mu\neq\pm$. Note that the correspondence is exact for the state insensitive to the phonons so that one defines a dressed state $|\chi_{o}\rangle\equiv |\psi_{o}\rangle$ ($\nu=L/2\equiv o$). By contrast, the correspondence disappears for the quasi-degenerate states and the smaller $\epsilon$ is, the larger is the energy difference. A strong hybridization occurs between  $|\psi_{\pm}\rangle$ that results in the formation of two quasi-degenerate dressed states. Denoted $|\chi_{\pm}\rangle$ ($\nu=L/2\mp1 \equiv \pm$), these dressed states mainly correspond to superimpositions involving $|\psi_{+}\rangle$ and $|\psi_{-}\rangle$.

\begin{figure}
\includegraphics{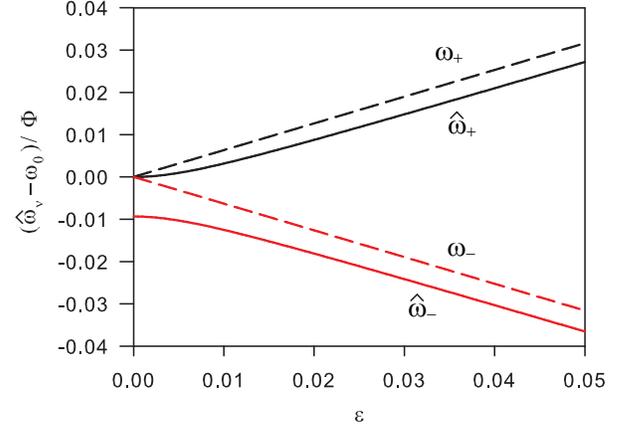}
\caption{(Color online) $\epsilon$ dependence of $\hat{\omega}_{\pm}$ (full lines) and 
$\omega_{\pm}$ (dashed lines) for $\chi=10$ pN and $L=10$.}
\end{figure}

The $\epsilon$ dependence of the energy $\hat{\omega}_{\pm}$ of the quasi-degenerate dressed state $|\chi_{\pm}\rangle$ is shown in Fig. 7. For quite large $\epsilon$ values, $\hat{\omega}_{+}$ (resp. $\hat{\omega}_{-}$) is slightly red-shifted when compared with 
$\omega_{+}$ (resp. $\omega_{-}$). In that case, $\hat{\omega}_{\pm}$ follows the $\epsilon$ dependence of the corrected bare energy $\omega_{\pm}+\delta \omega_{\pm}$ and $|\chi_{+}\rangle$ (resp. $|\chi_{-}\rangle$) basically corresponds to $|\psi_{+}\rangle$ (resp. $|\psi_{-}\rangle$). As $\epsilon$ decreases down to zero, $\hat{\omega}_{+}$ decreases and it reaches $\omega_0$. By contrast, $\hat{\omega}_{-}$ increases and it converges to a value located below the band center. The energy difference characterizes an anti-crossing phenomena so that the energy levels of the dressed states repel each other. This is the signature of the hybridization between $|\psi_{+}\rangle$ and $|\psi_{-}\rangle$ that results in the formation of two quasi-degenerate dressed states $|\chi_{\pm}\rangle \approx (|\psi_{+}\rangle\pm|\psi_{-}\rangle)/\sqrt{2}$.

\begin{figure}
\includegraphics{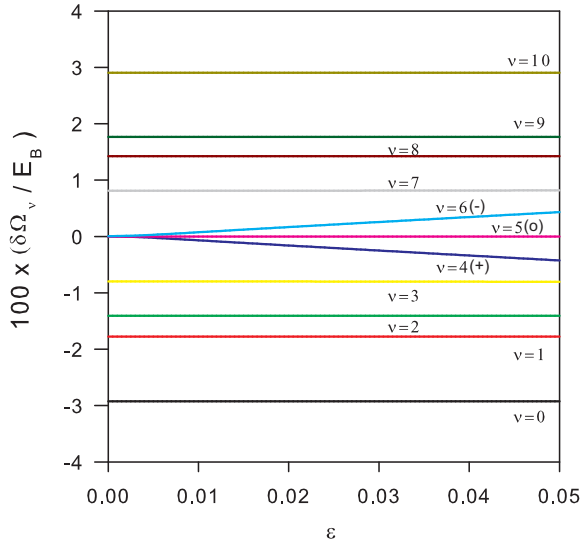}
\caption{(Color online) $\delta \Omega_{\nu}$ vs $\epsilon$ for $\chi=10$ pN and $L=10$.}
\end{figure}

The $\epsilon$ dependence of the phonon frequency correction $\delta \Omega_{\nu}$ is shown in Fig. 8. The phonon frequency is either red-shifted or blue-shifted depending on the nature of the exciton. 
A redshift is induced when the exciton occupies a dressed state $|\chi_{\nu}\rangle$ with $\nu=0,...,L/2-1$. The closer to the band edge the exciton energy is located, the larger is the shift. By contrast, an exciton that occupies a state $|\chi_{\nu}\rangle$ with $\nu=L/2+1,...,L$ yields a blueshift of the phonon frequency. As previously, excitonic states near the band edge favor the largest phonon frequency shift. 
Note that in the state $|\chi_{o}\rangle$, the exciton produces no phonon frequency shift. As shown in Fig. 8, when the exciton occupies a state isomorphic to a stationary wave, the phonon frequency shift is $\epsilon$ independent. By contrast, a strong $\epsilon$ dependence occurs when the exciton is in a quasi-degenerate dressed state $|\chi_{\pm}\rangle$. In that case, $\delta \Omega_{+}<0$ increases when $\epsilon$ decreases whereas $\delta \Omega_{-}>0$ decreases when $\epsilon$ decreases. For nonvanishing $\epsilon$ values, it turns out that $\delta \Omega_{-}\approx -\delta \Omega_{+}$. However, when $\epsilon$ tends to zero, $\delta \Omega_{+}$ converges to zero whereas $\delta \Omega_{-}$ reaches a quite small nonvanishing value. In other words, when $\epsilon \rightarrow 0$, the frequency shift is negligible when the phonon is accompanied by an exciton in either $|\chi_{o}\rangle$, $|\chi_{+}\rangle$ or $|\chi_{-}\rangle$.

\begin{figure}
\includegraphics{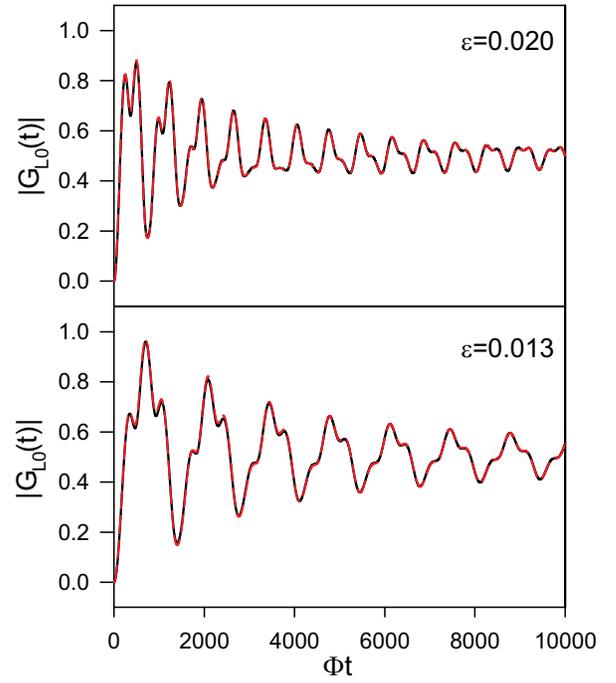}
\caption{(Color online) Time evolution of $|G_{L0}(t)|$ for $\chi=10$ pN, $L=10$ and $T=300$ K. Exact calculations (dashed lines) and PT (full lines).}
\end{figure}

The key ingredients entering PT being characterized, let us now study the effective exciton propagator $|G_{L0}(t)|$ whose time evolution is displayed in Fig. 9. The figure shows that PT perfectly agrees with exact calculations over a long-time scale. Initially equal to zero, $|G_{L0}(t)|$ first increases with time. Then, at time $T_{M}$, it reaches a maximum value $G_{M}$ quite close to unity. Finally, it develops damped oscillations that fluctuate around $1/2$. These oscillations support a high-frequency small-amplitude modulation whose behavior depends on $\epsilon$. When $\epsilon=0.020$, $|G_{L0}(t)|$ exhibits a double maximum. At $t=248.47\Phi^{-1}$, $|G_{L0}(t)|$ first reaches a local maximum whose value is equal to $0.83$. Then, the absolute maximum $G_{M}=0.88$ occurs at $T_{M}=495.48\Phi^{-1}$. Such a temporal structure is very sensitive to $\epsilon$. When $\epsilon$ slightly varies around $0.020$, the absolute maximum jumps between the two values of the double maximum so that $T_{M}$ exhibits a discontinuous character with respect to $\epsilon$. 
By contrast, when $\epsilon=0.013$, $|G_{L0}(t)|$ exhibits a single absolute maximum whose behavior remains quite stable when $\epsilon$ varies. It occurs at $T_{M}=699.85\Phi^{-1}$ and its value $G_{M}=0.97$ is very close to unity.

\begin{figure}
\includegraphics{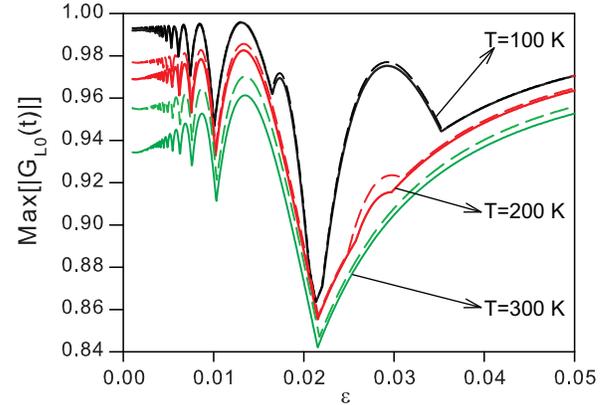}
\caption{(Color online) $Max[|G_{L0}(t)|]$ vs $\epsilon$ for $\chi=10$ pN and $L=10$. Exact calculations (dashed lines) and PT (full lines).}
\end{figure}

The $\epsilon$ dependence of the maximum value $G_M=Max[|G_{L0}(t)|]$ is illustrated in Fig. 10 for different temperatures. The figure shows that PT provides results in a quite good agreement with exact calculations. A small discrepancy occurs at high temperature and for small $\epsilon$ values, only. When $\epsilon$ decreases from $0.05$, the curve 
$G_M$ vs $\epsilon$ exhibits a series of minima and maxima whose value depends on the temperature. A local minimum corresponds to a singularity. The curve exhibits a kind of cusp in the neighborhood of which the time $T_M(\epsilon)$ is discontinuous. By contrast, a local maximum corresponds to a well defined point where the first derivative of $G_M(\epsilon)$ vanishes. Close to a local maximum, $T_M(\epsilon)$ remains continuous. 
Whatever the temperature, it turns out that the absolute minimum of $G_M(\epsilon)$ occurs for $\epsilon\approx0.021$. Its value slightly decreases with the temperature and it varies from $0.86$ for $T=100$ K to $0.84$ for $T=300$ K. By contrast, the absolute maximum takes place for $\epsilon\approx0.013$ $\forall$ $T$. Equal to $0.99$ for $T=100$ K, it decreases to $0.97$ for $T=300$ K. Note that if one defines $P=100\times(1-G_M)$ as the percentage of the lost information, one obtains $P\approx3$ \% at high temperature. 
When $\epsilon$ tends to zero, the maxima and the minima are becoming more frequent but less pronounced. $G_M$ tends to a temperature dependent value approximately equal to $0.99$ and $0.95$ for $T=100$ and $300$ K, respectively. We have verified that the curve $G_M$ vs $\epsilon$ depends on both the lattice size and the exciton-phonon coupling strength. These parameters modify the position and the value of the minima and of the maxima. For instance, for $\epsilon=0.01$ and $T=300$ K, one obtains $G_M=0.93$ for $L=10$ and $\chi=10$ pN, $G_M=0.89$ for $L=20$ and $\chi=10$ pN and $G_M=0.73$ for $L=10$ and $\chi=20$ pN. Increasing either $L$ or $\chi$ reduces the QST fidelity.

\begin{figure}
\includegraphics{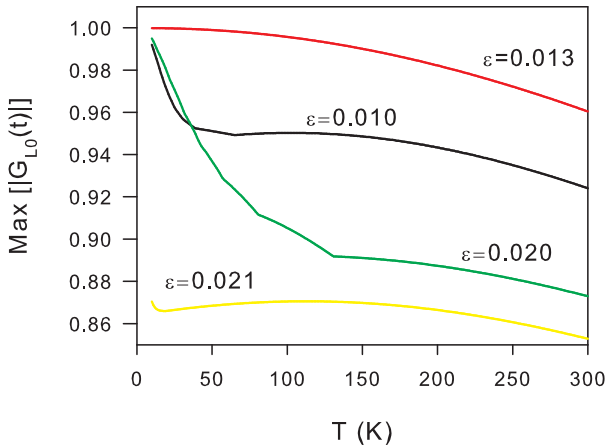}
\caption{(Color online) $Max[|G_{L0}(t)|]$ vs $T$ for $\chi=10$ pN and $L=10$. The calculations have been carried out using PT.}
\end{figure}

Finally, the temperature dependence of $G_M$ is illustrated in Fig. 11 for different $\epsilon$ values. The calculations have been carried out using PT that is particularly suitable for the considered $\epsilon$ values.  When $\epsilon=0.021$, the curve $G_M(\epsilon)$ lies in the neighborhood of its absolute minimum. Consequently, over the temperature range displayed in the figure, $G_M$ takes the smallest values. Almost uniform, it is approximately equal to $0.86\pm0.01$. The corresponding lost information is thus about $14$ \%. For $\epsilon=0.01$ and $0.02$, a critical temperature $T^{*}$ discriminates between two regimes. When $T<T^{*}$, $G_M$ rapidly decreases with the temperature. By contrast, when $T>T^{*}$, smooth variations take place. Note that $T^{*}\approx 40$ K for $\epsilon=0.01$ whereas $T^{*}\approx 130$ K for $\epsilon=0.02$. Finally, 
for $\epsilon=0.013$, the curve $G_M(\epsilon)$ lies in the neighborhood of its absolute maximum. Consequently,
$G_M$ slightly decreases with the temperature. It varies from (almost) unity for $T=10$ K to $0.97$ for $T=300$ K. It scales as $G_M\approx 1-(T/T_0)^2$ with $T_0\approx 1510$ K so that the lost information remains smaller than $3$ \%. 

\section{Discussion}

According to the numerical results, PT is a powerful tool for describing the exciton-phonon dynamics. Within the nonadiabatic weak-coupling limit, it accurately accounts for the spectral properties of the system over a broad energy scale. Moreover, PT is particularly suitable for characterizing the time evolution of the effective exciton propagator over a long-time scale, even at high temperature. In that context, both exact and approximate calculations have revealed the potential powerfulness of the proposed  communication protocol. The key point is that its efficiency strongly depends on the coupling $\epsilon$ between the QCs and the CC. When $\epsilon$ is judiciously chosen, it turns out that the maximum value of $|G_{L0}(t)|$ is quite close to unity. A high-fidelity QST occurs over a broad temperature range, the lost information during the transfer remaining smaller than 3 $\%$. By contrast, specific $\epsilon$ values induce a hole in the transmission curves. The maximum value of the effective exciton propagator deviates from unity resulting in the impoverishment of the transfered information. The lost information drastically increases and it can represent almost 15 \% of the initially implemented infromation at high temperature. Of course, these results depend on both the lattice size and the exciton-phonon coupling strength whose the increase reduces the fidelity of the transfer.
 
To interpret these results, PT can be used for deriving a simplified expression of $G_{L0}(t)$. Indeed, 
our numerical studies have revealed that the so-called diagonal approximation works quite-well. Consequently, the exact exciton-phonon eigenstates $|\Psi_{i}\rangle$ basically reduce to the effective Hamiltonian eigenstates $|\chi_{\nu}\rangle\otimes |n\rangle$. The transformation $U$ in Eq.(\ref{eq:GL01}) behaves as the unit operator and the exciton-phonon entanglement mainly results from the modification of the exciton states $|\psi_{\mu}\rangle \rightarrow |\chi_{\nu}\rangle$, the renormalization of the exciton energies $\omega_{\mu} \rightarrow \hat{\omega}_{\nu}$ and the correction of the phonon frequency $\Omega \rightarrow \Omega+\delta \Omega_{\nu}$. $G_{L0}(t)$ is thus rewritten as  
\begin{equation}
G_{L0}(t) \approx \sum_{\nu=0}^{L}  F_{\nu}(t) e^{-i\hat{\omega}_{\nu} t}  \langle L | \chi_{\nu} \rangle \langle \chi_{\nu} | 0 \rangle,
\label{eq:GL0APPROX}
\end{equation}
where $F_{\nu}(t)=Z_B^{(\nu)}(t)/Z_B$ is the decoherence factor as
\begin{equation}
F_{\nu}(t)=\frac{1-e^{-\beta \Omega}}{1-e^{-\beta \Omega-i\delta \Omega_{\nu} t}}.
\label{eq:Fnu1}
\end{equation}
In Eq.(\ref{eq:GL0APPROX}), $G_{L0}(t)$ is the sum of the probability amplitudes associated to the different paths that the exciton can follow to tunnel between the QCs. A given path defines a transition through the dressed state $|\chi_{\nu}\rangle$. The corresponding amplitude involves the weight of the localized states in the dressed state, a phase factor that accounts for the free evolution of the dressed exciton and the decoherence factor. This later contribution originates in the exciton-phonon interaction. Indeed, when Eq.(\ref{eq:GL0}) is developed in the phonon number state basis, $G_{L0}(t)$ can be interpreted as follows. The system being prepared in the factorized state $|0\rangle \otimes | n \rangle$,  $G_{L0}(t)$ is the probability amplitude to observe the system in a factorized state $|L\rangle \otimes \exp(-i n \Omega t) |n\rangle$. It thus describes an excitonic transition during which the phonons evolve freely. When the exciton occupies the state $|\chi_{\nu}\rangle$, the phonon frequency is modified. The probability amplitude that the phonons evolve freely reduces to a phase factor $\exp(-i n \delta \Omega_{\nu}t)$. Of course, such a phase factor does not affect the excitonic coherence when the phonons are initially in a pure state. However, at finite temperature, the phonons are described by a statistical mixture of number states. A thermal average is required and it yields a sum over different phase factors which interfere with the other. The interferences lead to a decay of the excitonic coherence encoded in the decoherence factor. 

At this step, Eq.(\ref{eq:GL0APPROX}) can be simplified because only the dressed states that involve the localized states $|0\rangle$ and $|L\rangle$ contribute significantly to the exciton propagator. Indeed, the bare states $|\psi_{\mu}\rangle$ that correspond to the stationary waves of the CC ($\mu\neq o,\pm$) are not degenerated. The dressing by virtual phonons mainly induces energy renormalization without significantly modifying their nature. Therefore, the system exhibits $N-1$ dressed states isomorphic to these stationary waves, i.e. $| \chi_{\mu} \rangle \approx |\psi_{\mu}\rangle$ and $\hat{\omega}_{\mu} \approx \omega_{\mu}+\delta \omega_{\mu}$ for $\mu\neq o,\pm$. Because these dressed states are almost independent on the localized states, their contribution to $G_{L0}(t)$ can be disregarded. Note that $|\psi_{\mu}\rangle$ is coupled with $z_{\mu}$ neighboring states through phonon exchanges. In the nonadiabatic limit ($|\omega_{\mu}-\omega_{\mu\pm1}|\ll \Omega$), Eq.(\ref{eq:deltaH}) yields $\delta \omega_{\mu}\approx -z_{\mu}E_B/L$, as observed in Fig. 5. 
Because $z_1=z_L=1$, the shift experienced by the band edge states $\mu=1$ and $\mu=L$ is two times smaller than the shift experienced by the remaining stationary waves for which $z_\mu=2$. In a market contrast, the bare state $|\psi_{o}\rangle$ does not interact with the phonons. The corresponding dressed state $|\chi_{o}\rangle \equiv |\psi_{o}\rangle$ only involves the localized states $|0\rangle$ and $|L\rangle$ (see Eq.(\ref{eq:STATE1})) and the decoherence factor $F_{o}(t)$ reduces to unity. Consequently, $|\chi_{o}\rangle$ defines an ideal path for the excitonic transition between the QCs. 

The fundamental point concerns the quasi-degenerate states $|\psi_{\pm}\rangle$ that are profoundly perturbed by the dressing mechanism. As for the stationary waves, the exciton-phonon interaction induces a redshift $\delta\omega_{\pm}$ of each energy $\omega_{\pm}$. This interaction originates in the dependence of $|\psi_{\pm}\rangle$ with respect to the robust stationary wave $|\varphi_{L/2}\rangle$. Therefore, each quasi-degenerate state is coupled with two stationary waves through phonon exchanges. In that case, it is easy to extract $\delta \omega_{\pm}$ from Eq.(\ref{eq:deltaH}). In doing so, the shifts can be expanded as a Taylor series with respect to $\epsilon$ as $\delta \omega_{\pm}=-\sum_{r=0}^{\infty} (-1)^rE_r\epsilon^r$. The positive coefficients $E_r$ are defined as 
\begin{equation}
E_{r}=\frac{\eta^2}{2} \left( \frac{\bar{E}^r}{(\Omega+\Delta\omega)^{r+1}}+ \frac{\bar{E}^r}{(\Omega-\Delta \omega)^{r+1}} \right),
\label{eq:Er}
\end{equation}
where $\bar{E}=2\Phi/\sqrt{L}$ and $\Delta \omega = |\omega_{L/2}^0-\omega_{L/2\pm1}^0|$. 
In the nonadiabatic limit, $E_0\approx -E_B/L$ so that $\delta\omega_{+}$ (resp. $\delta\omega_{-}$)  decreases (resp. increases) linearly with $\epsilon$ form $-E_B/L$, in a quite good agreement with the results displayed in Fig. 5. Note that  
$\delta\omega_{\pm}$ is similar to the shift experienced by the band edge states because the weight
of $|\varphi_{L/2}\rangle$ in $|\psi_{\pm}\rangle$ is equal to $1/\sqrt{2}$.

In addition to energy renormalization, the exciton-phonon interaction yields an effective coupling $v_{+-}=-(\delta \omega_{+}+\delta \omega_{-})/2$ between $|\psi_{+}\rangle$ and $|\psi_{-}\rangle$ (Eq.(\ref{eq:deltaH})). Quite similar in magnitude to the energy shifts, this coupling can no longer be neglected resulting in a strong hybridization between the quasi-degenerate states. These states behave as a two-level system independent on the remaining states whose eigenstates define two quasi-degenerate dressed states as
\begin{eqnarray}
|\chi_{+}\rangle&\approx&+\cos(\theta)|\psi_{+}\rangle + \sin(\theta)|\psi_{-}\rangle \nonumber \\
|\chi_{-}\rangle&\approx&-\sin(\theta)|\psi_{+}\rangle + \cos(\theta)|\psi_{-}\rangle.
\label{eq:STATECHI}
\end{eqnarray}
The corresponding eigenenergies are expressed as 
\begin{equation}
\hat{\omega}_{\pm}=\omega_0+\frac{\delta \omega_{+}+\delta \omega_{-}}{2}\pm \frac{1}{2}\sqrt{\Delta},
\label{eq:ECHI}
\end{equation}
where
\begin{eqnarray}
\Delta&=&(\omega_{+}+\delta \omega_{+}-\omega_{-}-\delta \omega_{-})^2+4v_{+-}^2 \nonumber \\
\cos(2\theta)&=&\frac{\omega_{+}+\delta \omega_{+}-\omega_{-}-\delta \omega_{-}}{\sqrt{\Delta}}.
\label{eq:theta}
\end{eqnarray}
The hybridization is encoded in the $\theta$ parameter whose value depends on both the exciton-phonon coupling strength, measured by $\delta \omega_{\pm}$, and the energy difference $\delta=\omega_{+}-\omega_{-}\equiv2\bar{E}\epsilon$. Two asymptotic situations occur. For quite large $\epsilon$ values, $|\delta \omega_{\pm}| \ll \delta$ so that the influence of the quasi-degeneracy is negligible. The hybridization is weak and the dressed states reduce to the bare states, i.e. $|\chi_{\pm}\rangle\approx |\psi_{\pm}\rangle$ and $\hat{\omega}_{\pm}\approx \omega_{\pm}+\delta \omega_{\pm}$. The bare energies experience a similar redshift approximately equal to $-E_B/L$, as displayed in Fig. 7 for $\epsilon\approx 0.05$. 
By contrast, for small $\epsilon$ values, $|\delta \omega_{\pm}| \gg \delta$ and the hybridization is strong. One obtains $\theta\approx \pi/4$ and $|\chi_{\pm}\rangle\approx (|\psi_{+}\rangle\pm|\psi_{-}\rangle)/\sqrt{2}$. 
The energy $\hat{\omega}_{+}$ reaches the band center and it scales as $\hat{\omega}_{+}\approx \omega_0+(\bar{E}-E_1)^2\epsilon^2/2E_0$. By contrast, $\hat{\omega}_{-}\approx \omega_0-2E_0-2E_2\epsilon^2-(\bar{E}-E_1)^2\epsilon^2/2E_0$ lies close to $\omega_0-2E_B/L$, in a good agreement with the results displayed in Fig. 7. Note that in that case $|\chi_{+}\rangle$ reduces to a superimposition of the localized states whereas $|\chi_{-}\rangle$ tends to the robust stationary wave. 

By combining Eqs.(\ref{eq:STATE1}) and (\ref{eq:STATECHI}), it is easy to show that the quasi-degenerate dressed states depend on the localized states. They thus define two relevant paths for the excitonic transition between the QCs. However, when the exciton occupies $|\chi_{\pm}\rangle$, it induces a shift $\delta \Omega_{\pm}$ of the phonon frequency. Quantum decoherence is no longer negligible and the decoherence factors is expressed as 
\begin{equation}
F_{\pm}(t)\approx \frac{e^{-i\bar{n} \delta \Omega_{\pm}t}}{\sqrt{1+4\Delta \bar{n}^2 \sin^2(\delta \Omega_{\pm}t/2)}},
\label{eq:Fnu2}
\end{equation}
where $\bar{n}=[\exp(\beta \Omega)-1]^{-1}$ and $\Delta \bar{n}^2=\bar{n}(\bar{n}+1)$. Note that Eq.(\ref{eq:Fnu2}) has been obtained from Eq.(\ref{eq:Fnu1}) by simplifying the time dependence of the argument of the decoherence factor. By inserting  Eq.(\ref{eq:STATECHI}) into Eq.(\ref{eq:deltaH}), the phonon frequency shift is written as 
\begin{equation}
\delta \Omega_{\pm} \approx \mp 2\cos(2\theta)\sum_{r=0}^{\infty}E_{2r+1}\epsilon^{2r+1}.
\label{eq:dOM}
\end{equation}
In a quite good agreement with the results displayed in Fig. 8, Eq.(\ref{eq:dOM}) reveals that when the exciton occupies the state $|\chi_{+}\rangle$ (resp. $|\chi_{-}\rangle$), it induces a redshift (resp. blueshift) of the phonon frequency. Moreover, as observed in Fig. 8 when $\epsilon$ is not too small, one obtains $\delta \Omega_{-}=-\delta \Omega_{+}$. However, when $\epsilon$ tends to zero, Eq.(\ref{eq:dOM}) yields $\delta \Omega_{\pm}\approx \mp2E_1(\bar{E}-E_1)\epsilon^2/E_0$. Such a behavior slightly differs from our numerical results. As shown in Fig. 8, $\delta \Omega_{+}$ vanishes when $\epsilon$ tends to zero whereas $\delta \Omega_{-}$ reaches a rather small value. 
This discrepancy originates in the fact that the exact dressed states $|\chi_{\pm}\rangle$ slightly depend on the stationary waves $|\psi_{\mu}\rangle$, with $\mu\neq o,\pm$. When $\epsilon$ tends to zero, such a dependence is more pronounced for $|\chi_{-}\rangle$. These additional components, that are not taken into account in Eq.(\ref{eq:STATECHI}), yield a nonvanishing $\delta \Omega_{-}$ value. 

In that context, it turns out that $|\chi_{o}\rangle$, $|\chi_{+}\rangle$ and $|\chi_{-}\rangle$ 
define the main paths followed by the exciton to tunnel between the QCs. Consequently, $G_{L0}(t)$ is rewritten as
\begin{eqnarray}
G_{L0}(t) &\approx& -\Delta_N \exp(-i\omega_{0} t) \left[ +\frac{1}{2} \right. \label{eq:GL0approx}    
\\
          &-&\frac{1}{4}|F_{+}(t)| \exp(-iW_{+} t) [1+\sin(2\theta)] \nonumber \\
          &-& \left. \frac{1}{4}|F_{-}(t)| \exp(+iW_{-} t) [1-\sin(2\theta)] \right],  \nonumber
\end{eqnarray}
where $W_{\pm}=\pm(\hat{\omega}_{\pm}-\omega_0+\bar{n}\delta \Omega_{\pm})$ defines a positive frequency relative to the band center $\omega_0$. According to Eq.(\ref{eq:GL0approx}), $G_{L0}(t)$ is the sum of the three transition amplitudes, $\tau_o=1/2$ and $\tau_{\pm}=-|F_{\pm}(t)| \exp(\mp iW_{\pm} t) [1\pm\sin(2\theta)]/4$, connected to the three relevant paths. Because $|\chi_{o}\rangle$ is insensitive to the phonons, the probability amplitude $\tau_o$ is the weight $1/2$ of the localized states. By contrast, the probability amplitude $\tau_{\pm}$ that the exciton tunnels through $|\chi_{\pm}\rangle$ involves a phase factor whose time evolution is governed by $T_{\pm}=\pi/W_{\pm}$. This phase factor is weighted by both the decoherence factor and the weight of the localized states. 
Consequently, depending on the value of the model parameters, $W_{\pm}$, $\theta$ and $F_{\pm}(t)$ will take particular values so that each probability amplitude will develop a specific time evolution. Quantum interferences between the different amplitudes will occur resulting in a characteristic time evolution of $|G_{L0}(t)|$. As time evolves, $|G_{L0}(t)|$ will reach a maximum whose value $G_M$ will depend on the model parameters through the interference pattern.    

\begin{figure}
\includegraphics{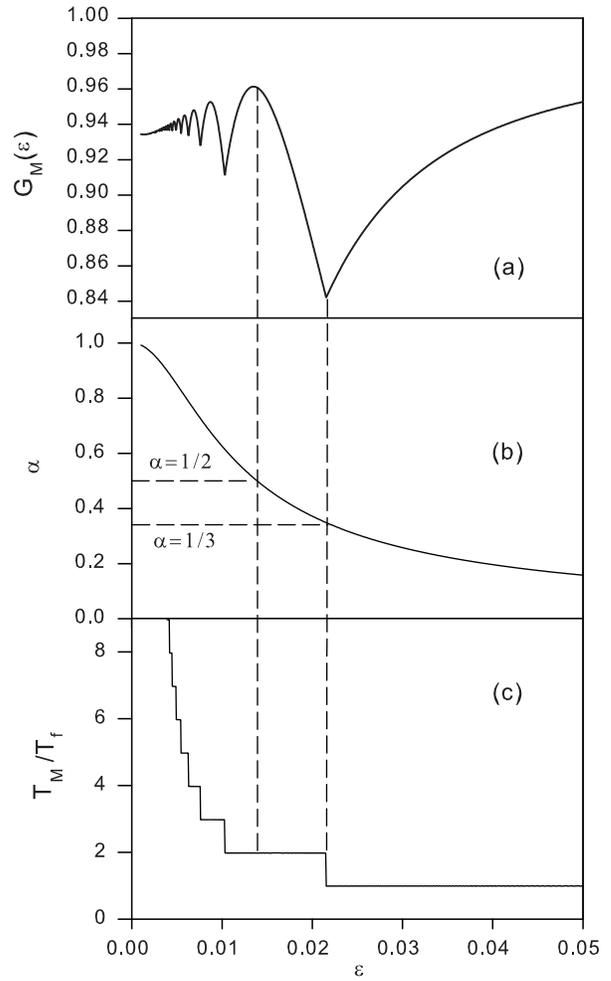}
\caption{(a) $G_M(\epsilon)$ vs $\epsilon$ for $\chi=10$ pN, $L=10$ and $T=300$ K using PT. (b) $\alpha=W_s/W_f$ vs $\epsilon$. (c) $T_M/T_f$ vs $\epsilon$.}
\end{figure}

To understand this phenomenon intuitively, let us focus our attention on the influence of the coupling $\epsilon$ between the QCs and the CC. Provided that $\epsilon$ is not too small, the numerical results suggest that $\delta \Omega_{-}\approx-\delta \Omega_{+}>0$. The modulus of the decoherence factor is thus almost independent on the nature of the exciton so that  $|F_{\nu}(t)| \approx F(t)$ $\forall \nu=\pm$. Moreover, it turns out that the influence of $\theta$ remains negligible, i.e.  $\sin(2\theta)\approx 0$. Therefore, after straightforward algebraic manipulations, one obtains
\begin{eqnarray}
|G_{L0}(t)|^2 \approx \frac{1}{4} [1 &+& F^2(t) \cos^2(W_s t)  \label{eq:Gmod1}    \\
          &-& 2F(t)\cos(W_s t)\cos(W_f t)  ],  \nonumber
\end{eqnarray}
where $W_{s}=(W_{-}-W_{+})/2$ and $W_{f}=(W_{-}+W_{+})/2$. In the right hand side of Eq.(\ref{eq:Gmod1}), the first term  is the probability $|\tau_{o}|^2$ that the exciton realizes a transition through $|\chi_{o}\rangle$. The second term reduces to $|\tau_{+}+\tau_{-}|^2$ and it refers to the probability that the exciton tunnels through either $|\chi_{+}\rangle$ or $|\chi_{-}\rangle$. The last term describes the quantum interferences that arise when the exciton follows the paths involving either $|\chi_{0}\rangle$ or $|\chi_{+}\rangle$, and $|\chi_{0}\rangle$ or $|\chi_{-}\rangle$. It mixes $W_+$ and $W_-$ and it behaves as an amplitude-modulated signal $f(t)=F(t)\cos(W_s t)\cos(W_f t)$. It thus exhibits a slowly varying envelope with frequency $W_s$ that supports a fast modulation with frequency $W_f$, both components being damped by the decoherence factor $F(t)$. Such a behavior allows us us to introduce the fundamental times $T_f=\pi/W_f$ and $T_s=\pi/W_s$.

The time evolution of $|G_{L0}(t)|$ is mainly encoded in the quantum interference term $f(t)$. Consequently, $|G_{L0}(t)|$ reaches its maximum value $G_M(\epsilon)$ for a time $T_M(\epsilon)$ that minimizes $f(t)$. As illustrated in Fig. 12(c), such a situation occurs when $T_M$ is a multiple of $T_f$, i.e. $T_M=nT_f$ with $n=1,2,...$. However, the integer $n$ and the magnitude of $G_M(\epsilon)$ depend on $\epsilon$ through the value of the ratio $\alpha=W_s/W_f$. This ratio extends from zero for large $\epsilon$ values to unity for small $\epsilon$ values, as displayed  in Fig. 12(b). Therefore, as $\epsilon$ varies, different regimes take place.

For specific $\epsilon$ values, $\alpha$ is such that a situation arises in which the quantum interference term $f(t)$ exhibits two equivalent minima at two distinct times. In that case, the curve $G_M$ vs $\epsilon$ exhibits a local minimum and the time $T_M$ shows a discontinuity between two multiples of $T_f$. We have observed that in that case there are two equivalent configurations for which the probability amplitudes $\tau_+$ and $\tau_-$ are in phase with each other but are not exactly in phase with the amplitude $\tau_o$. At high temperature, such a situation occurs for the $\epsilon$ values that provide $\alpha=(2q-1)/(2q+1)$ with $q=1,2,...$. The time $T_M$ exhibits discontinuities between $qT_f$ and $(q+1)T_f$ and one obtains $T_+=2qT_-$ and $T_f=2T_+/(2q+1)$. Note that as shown in Fig. 12, the largest hole in the transmission curve occurs for $q=1$, i.e. for $\alpha=1/3$, so that $T_M$ varies between $T_f$ and $2T_f$.

In a marked contrast, for particular $\epsilon$ values, $\alpha$ is such that the quantum interference term $f(t)$ exhibits an isolated minimum close to $-1$. Such a situation takes place at a time $T_M(\epsilon)$ so that the probability amplitudes $\tau_+$ and $\tau_-$ are simultaneously in phase with the probability amplitude $\tau_o$. Constructive interferences occur between the different paths followed by the exciton to tunnel between the QCs. As a result, the curve $G_M$ vs $\epsilon$ exhibits a local maximum (see Figs. 10 and 12(a)). Such a situation appears for the $\epsilon$ values
that yield $\alpha=(q-p)/(q+p+1)$, where $p$ and $q$ are two positive integers such that $q>p$. One thus obtains $T_M=(2p+1)T_+=(2q+1)T_-$, i.e. $T_M=(p+q+1)T_f$. In that case, $f(T_M)=-F(T_M)$ so that the maximum value of the exciton effective propagator is approximately equal to $G_M=[1 + F(T_M)]/2$.

As shown in Figs. 10 and 12(a), the largest local maximum occurs for a specific value $\epsilon^*$. When 
$\epsilon=\epsilon^*$, $p=0$ and $q=1$ so that $\alpha=1/2$ and $T_M=T_+=3T_-$. Solving the equation $T_{+}=3T_{-}$ by exanding the parameters with respect to $\epsilon$ provides an analytical expression for $\epsilon^*$ written as 
\begin{equation}
\epsilon^*\approx \frac{\sqrt{2}E_0}{\sqrt{(\bar{E}-E_1)^2-4\bar{n}E_1(\bar{E}-E_1)-2E_0E_2 }}
\label{eq:optim1}
\end{equation}
Over the temperature range $T=100$ - $300$ K and for $\chi=10$ pN and $L=10$, Eq.(\ref{eq:optim1}) yields $\epsilon^*\approx 0.011$ in a quite good agreement with the numerical estimate $\epsilon^*\approx 0.013$ (Fig. 10). 
Therefore, when $\epsilon=\epsilon^*$, $T_+(\epsilon^*)$ defines the shortest time for which constructive interferences occur. The decay provided by the decoherence factor is thus minimized and the QST is optimized. In the nonadiabatic weak-coupling limit one thus obtains 
\begin{equation}
G_M(\epsilon^*)\approx \left[ 1-\frac{\pi^2}{4} \frac{\Delta \bar{n}^2}{\delta n^2} \right]
\label{eq:optim}
\end{equation}
where $\delta n=n_0-\bar{n}$ and $n_0=|\hat{\omega}_{+}(\epsilon^*)-\omega_0| / |\delta \Omega_{+}(\epsilon^*)|$.
Eq.(\ref{eq:optim}) clearly shows that the impoverishment of the transfered information results from the thermal fluctuations $\Delta \bar{n}$ of the phonon number. Fortunately, in the weak coupling limit, $n_0\gg\bar{n}$ so that $\Delta \bar{n}$ is always smaller than $\delta n$. Because $\Delta \bar{n}$ is proportional to $k_BT/\Omega$, the optimized value of the effective exciton propagator scales as $G_M(\epsilon^*)\approx  1-(T/T_0)^2$ with $T_0=2n_0\Omega/\pi$. As observed in Fig. 11, $G_M(\epsilon^*)$ only slightly deviates from unity provided that $T$ remains smaller than the critical temperature $T_0$. With the parameters used in the simulation ($\chi=10$ pN and $L=10$), $T_0\approx 1600$ K indicating that the lost information during the transfer is negligible, even at room temperature. 

Consequently, when $\epsilon$ is judiciously chosen, constructive interferences take place between the different paths followed by the exciton to tunnel between the QCs. The influence of the quantum decoherence is minimized and an ideal QST occurs over a broad temperature range. The fidelity of the transfer remains quite close to unity, even at high temperature, indicating the powerfulness of the proposed communication protocol.

\section{Conclusion}

In the present paper, a new communication protocol has been proposed in which QST is achieved by a high-frequency vibrational exciton. The main idea was to use two distant molecular groups grafted on the sides of a molecular lattice. These groups behave as two QCs on which the information is encoded and received. By contrast, the lattice defines the CC along which the exciton propagates and interacts with a phonon bath. To minimize the impact of the quantum decoherence, the structure was designed so that a vibrational resonance occurs between the QCs and the robust stationary wave of the lattice whose energy is exactly located at the band center.  

To highlight the relevance of PT, special attention has been paid for describing a simple model in which an exciton is dressed by a single phonon mode, only. In that case, the Hamiltonian was solved exactly so that the PT accuracy has been checked. Within the nonadiabatic weak-coupling limit, it has been shown that PT is a powerful tool for characterizing the exciton-phonon dynamics. Provided that the lattice size is not too large, it yields a very good estimate of the spectral properties over a broad energy scale. Moreover, it has been observed that PT is particularly suitable for describing the time evolution of the exciton RDM, even at high temperature. 

In that context, it has been shown that the system supports three quasi-degenerate exciton states that
involve the states localized on the QCs. When the exciton occupies one of these states, it does not significantly modify the phonon bath and keeps its coherent nature over a long-time scale. These states define the relevant paths followed by the exciton to tunnel between the QCs. Consequently, when the coupling between the QCs and the CC is judiciously chosen, constructive interferences take place between these paths. The quantum decoherence is minimized and an almost ideal QST occurs. The fidelity of the transfer remains quite close to unity over a broad temperature range, indicating the powerfulness of the proposed communication protocol. 

Finally, because the present approach has clearly revealed the relevance of PT, it will be generalized for describing the influence of all the phonon modes. The main idea is to check the efficiency of the proposed protocol with a more realistic system whose dynamics cannot be solved exactly.

\appendix 

\section{Quasi-degenerate second order perturbation theory}

Quasi-degenerate PT involves a unitary transformation $U=\exp(S)$ that provides a block-diagonal transformed Hamiltonian $\hat{H}=UHU^{\dag}$ in the unperturbed basis $| \Psi_{\mu,n}^{0}\rangle$. More precisely, the desired Hamiltonian must be diagonal with respect to the phonon number states, only. To proceed, any operator $O$ acting in $\mathcal{E}$ can be split as $O=O_D+O_{ND}$, where $O_D$ is the diagonal part with respect to the phonon number states whereas $O_{ND}$ is the remaining non-diagonal part. Note that such a partition is equivalent to that provided by a projector formalism\cite{kn:cohen}. In the unperturbed basis, these operators are defined as
\begin{eqnarray}
\langle \Psi_{\mu,n}^{0} | O_D | \Psi_{\mu',n'}^{0} \rangle &=& \langle \Psi_{\mu,n}^{0} | O | \Psi_{\mu',n}^{0} \rangle \delta_{nn'} \\
\langle \Psi_{\mu,n}^{0} | O_{ND} | \Psi_{\mu',n'}^{0} \rangle &=& \langle \Psi_{\mu,n}^{0} | O | \Psi_{\mu',n'}^{0} \rangle (1-\delta_{nn'}). \nonumber
\end{eqnarray}
In that context, because $V_D=0$, one seeks the anti-hermitian generator $S\equiv S_{ND}$ as a non-diagonal operator with respect to the phonon number states. It is expanded as a Taylor series as $S=S_1+S_2+...$ where $S_{q}$ is the $q$th order correction in the coupling $V$. Consequently, $\hat{H}$ becomes
\begin{eqnarray}
\hat{H}&=&H_0+V+[S_1,H_0] \nonumber \\
&+&[S_1,V]+[S_2,H_0]+\frac{1}{2}[S_1,[S_1,H_0]]+...
\label{eq:Ht}
\end{eqnarray}
From Eq.(\ref{eq:Ht}), $S$ is derived order by order to obtain a block-diagonal form for $\hat{H}$ at the desired order. 
Up to second order, the solution is given by the equations 
\begin{eqnarray}
&&\left[ H_0,S_1 \right] = V_{ND} \nonumber \\
&&\left[ H_0,S_2 \right] = \frac{1}{2} [ S_1,V ]_{ND}  \nonumber \\ 
&&\hat{H} = H_0+\frac{1}{2}[ S_1,V]_{D}.
\label{eq:S1S2}
\end{eqnarray}

Because $V=M(a^{\dag}+a)$ is a linear combination of creation and annihilation phonon operators, $S_1$ is of the form  
$S_1=Za^{\dag}-Z^{\dag}a$. The unknown operator $Z$ acts in $\mathcal{E}_A$, only. No restriction affects this operator because $S_1^{\dag}=-S_1$. Therefore, inserting this expression into Eq.(\ref{eq:S1S2}) yields 
$Z_{\mu\mu'} =M_{\mu \mu'}/(\omega_{\mu}-\omega_{\mu'}+\Omega)$.

The knowledge of $S_1$ allows us to compute the commutator $[S_1,V]$ that is required to derive both $\hat{H}$ and $S_2$. 
This commutator is defined as  
\begin{equation}
\frac{1}{2}[S_1,V]=A +Ba^{\dag}a^{\dag}+B^{\dag}aa+(B+B^{\dag})a^{\dag}a, 
\label{eq:S1V}    
\end{equation}
where $B= [Z,M]/2$ and $A=-(Z^{\dag}M+MZ)/2$. From the diagonal part of Eq.(\ref{eq:S1V}), $\hat{H}$ reduces to  
\begin{equation}
\hat{H}=H_A+A+ ( \Omega+B+B^{\dag}) a^{\dag}a.
\label{eq:Hx1}
\end{equation}
We thus recover Eq.(\ref{eq:HEFF}) with $\delta H_A=A$ and $\delta \Omega=B+B^{\dag}$ whose representation in the unperturbed basis yields Eq.(\ref{eq:deltaH}). From the non diagonal part of Eq.(\ref{eq:S1V}), one seeks $S_2$ of the form
$S_2=Ea^{\dag}a^{\dag}-E^{\dag} aa$. The unknown operator $E$ acts in $\mathcal{E}_A$, only. Inserting this expression into Eq.(\ref{eq:S1S2}) yields $E_{\mu \mu'}=B_{\mu \mu'}/(\omega_{\mu}-\omega_{\mu'}+2\Omega)$.

\section{Effective exciton propagator}

Expanding $U$ as a Taylor series with respect to $V$, Eq.(\ref{eq:GL01}) yields the second order expression of the effective exciton propagator as 
\begin{eqnarray}
G_{L0}(t)&=&\sum_{\nu=0}^{L}  \frac{Z_B^{(\nu)}(t)}{Z_B} e^{-i\hat{\omega}_{\nu} t}  \times \nonumber \\
&[&\langle L | \chi_{\nu} \rangle \langle \chi_{\nu} | 0 \rangle \nonumber \\
&+&\langle L |Z| \chi_{\nu} \rangle \langle \chi_{\nu} |Z^{\dag}| 0 \rangle n^{(\nu)}(t) e^{i(\Omega+\delta \Omega_{\nu})t} \nonumber \\
&+&\langle L |Z^{\dag}| \chi_{\nu} \rangle \langle \chi_{\nu} |Z| 0 \rangle (n^{(\nu)}(t)+1) e^{-i(\Omega+\delta \Omega_{\nu})t} \nonumber \\
&-&\langle L |ZZ^{\dag}| \chi_{\nu} \rangle \langle \chi_{\nu} | 0 \rangle n^{(\nu)}(t)/2 \nonumber \\
&-&\langle L |Z^{\dag}Z| \chi_{\nu} \rangle \langle \chi_{\nu} | 0 \rangle (n^{(\nu)}(t)+1)/2 \nonumber \\
&-&\langle L | \chi_{\nu} \rangle \langle \chi_{\nu} |ZZ^{\dag}| 0 \rangle n^{(\nu)}(t)/2 \nonumber \\
&-&\langle L | \chi_{\nu} \rangle \langle \chi_{\nu} |Z^{\dag}Z| 0 \rangle (n^{(\nu)}(t)+1)/2 \ \ ], 
\label{eq:GL0PT}
\end{eqnarray}
where $n^{(\nu)}(t)=[\exp(\beta \Omega +i\delta \Omega_{\nu} t)-1]^{-1}$.

\end{document}